# The optical afterglow of the short γ-ray burst GRB 050709

Jens Hjorth[1], Darach Watson[1], Johan P. U. Fynbo[1], Paul A. Price[2], Brian L. Jensen[1], Uffe G. Jørgensen[1], Daniel Kubas[3], Javier Gorosabel[4], Páll Jakobsson[1], Jesper Sollerman[1,5], Kristian Pedersen[1] & Chryssa Kouveliotou[6]

[1]*Dark Cosmology Centre, Niels Bohr Institute, University of Copenhagen, Juliane Maries Vej, DK-2100 Copenhagen Ø, Denmark.* [2]*Institute for Astronomy, University of Hawaii, 2680 Woodlawn Drive, Honolulu, Hawaii 96822, USA.* [3]*ESO Santiago, Casilla 19001, Santiago 19, Chile.* [4]*Instituto de Astrofisica de Andalucía (IAA-CSIC), PO Box 3004, E-18080 Granada, Spain.* [5]*Department of Astronomy, Stockholm University, AlbaNova, 106 91 Stockholm, Sweden.* [6]*NASA/Marshall Space Flight Center, National Space Science Technology Center, XD-12, 320 Sparkman Drive, Huntsville, Alabama 35805, USA.*

**It has long been known that there are two classes[1] of γ-ray bursts (GRBs), mainly distinguished by their durations. The breakthrough in our understanding of long-duration GRBs (those lasting more than ~2 s), which ultimately linked them with energetic Type Ic supernovae[2–4], came from the discovery of their long-lived X-ray[5] and optical[6,7] 'afterglows', when precise and rapid localizations of the sources could finally be obtained. X-ray localizations have recently become available[8,9] for short (duration <2 s) GRBs, which have evaded optical detection for more than 30 years. Here we report the first discovery of transient optical emission (R-band magnitude ~23) associated with a short burst; GRB 050709. The optical afterglow was localized with subarcsecond accuracy, and lies in the outskirts of a blue dwarf galaxy. The optical and X-ray[10] afterglow properties 34 h after the GRB are reminiscent of the afterglows of long GRBs, which are attributable to synchrotron emission from ultrarelativistic ejecta. We did not, however, detect a supernova, as found in most nearby long GRB afterglows, which suggests a different origin for the short GRBs.**

NASA's High Energy Transient Explorer, HETE-2, detected GRB 050709 at 22:36:37 UT on 9 July 2005 (ref. 9). The burst consisted of a single hard γ-ray pulse only 70 ms long, with peak energy of 83 keV, followed ~30 s later by fainter and softer emission from the same location, detected only at energies below 10 keV and lasting for about 2 min. This later component, speculated to be due to afterglow emission[9], allowed HETE-2's Soft X-ray Camera (SXC) to obtain a source location with an 81-arcsec error radius.

The properties of GRB 050709 place it firmly in the elusive short GRB population, as observed[1] by the Burst And Transient Source Experiment (BATSE). The first, hard pulse was shorter than most BATSE short-hard GRBs and had a peak energy and spectral slope consistent with the short-hard distribution[1,9,11] (Fig. 1). However, the second pulse of faint, soft X-ray emission, observed only by SXC, would not have been detectable with the BATSE Large Area Detectors, which had a low energy threshold of ~30 keV. Thus the entire event would have been classified as a short GRB in the BATSE sample. Moreover, we note that statistical studies that summed the background-subtracted signals of large numbers of BATSE bursts have also shown long-lasting (hundreds of seconds) fainter emission following the main event[12,13].

We observed the SXC error circle of GRB 050709 with the Danish 1.5-m telescope at the La Silla Observatory starting 33 h after the event, and obtained images in good observing conditions over the following 18 days (Table 1). We used point-spread-function-matched image subtraction[14] and established the existence of a single transient source inside the error circle positionally coincident with an X-ray source[10] observed with the Chandra X-ray Observatory (CXO) (Fig. 2). The point source is positioned at right ascension $\alpha_{2000}$=23 h 01 min 26.957 s, declination $\delta_{2000}$=−38° 58′ 39.76″, with an error of 0.25 arcsec, and sits on the edge of an extended source, apparently the host galaxy of the burst. Moreover, the source faded with a decay index ($f_\nu \propto t^\alpha$, where $f_\nu$ is the flux density and $t$ is time) of $\alpha$=−1.33±0.45 (Fig. 3), similar to the behaviour of afterglow emission from long GRBs. We therefore conclude that this source is the optical counterpart to the short GRB 050709. The extended source is an emission-line galaxy at a redshift $z$=0.16 (ref. 10). Assuming a flat Universe with Hubble constant $H_0$=72 km s$^{-1}$ Mpc$^{-1}$, matter density parameter $\Omega_m$=0.27 and equation of state parameter $w$=−1 gives a luminosity distance $d_L$=747 Mpc and a projected distance of the transient from the galaxy centre of ~3.5 kpc (1.3±0.1 arcsec), strongly suggesting that this is the host galaxy of the GRB. At this distance, the fluence of the short pulse converts to an isotropic-equivalent energy release of 2.2×10$^{49}$ erg (ref. 9). Additional multicolour imaging obtained at the Danish 1.5-m telescope (Table 1) shows that the host is a fairly blue dwarf galaxy with B-, V-, R- and I-band magnitudes of 22.22±0.10, 21.22±0.10, 21.24±0.12 and 20.38±0.25, respectively (uncorrected for the small Galactic extinction). At $z$=0.16 these colours indicate that the host is a starburst galaxy with little extinction, an age of 0.4 Gyr, a total luminosity of approximately 0.03$L^*$ (where $L^*$ is the characteristic luminosity of the Schechter luminosity function) (blue absolute magnitude $M_B$=−16.9±0.1), and a star formation rate of the order of 0.1$M_\odot$yr$^{-1}$ (where $M_\odot$ is the solar mass).

Our observations thus establish that the short-duration GRB 050709 was accompanied by long-lived optical afterglow emission, and most probably occurred in the outskirts of a star-forming dwarf galaxy. In these respects, GRB 050709 shares the properties of long-duration GRBs. However, several other properties, in addition to its short duration, set this event apart from long GRBs: The prompt emission[9] is less luminous than most long GRBs. In particular, GRB 050709 does not lie close to the luminosity–peak energy correlation established for long bursts[15]—its peak energy (~83 keV)[9] is much higher than observed in the few long GRBs that are this faint. Likewise, the X-ray afterglow[10] was about 1,000 times fainter than the equivalent isotropic X-ray luminosities of long GRB afterglows[16].

What do our optical observations tell us about the progenitor of GRB 050709? Short GRB progenitor models[17] include the merger of two compact objects, such as two neutron stars[18] or a neutron star and a black hole, or, in the 'unification scheme'[19,20], a variant of the collapsar model.

Our upper limits on optical emission between 7 and 20 days after the burst (Table 1) rule out the occurrence of an energetic Type Ic supernova (SN) as found in most long GRBs at low redshifts. To accommodate the non-detection in our data (which corresponds to an absolute magnitude of $M_V > -14.7$ at 16 days in the restframe), any supernova must have been at least 7 times fainter than the faint Type Ic SN 1994I (Fig. 3). A supernova similar to the canonical GRB 980425/SN 1998bw[2] would have reached a brightness of $R = 19.6$ mag, as was also observed in the long GRB 030329/SN 2003dh at $z = 0.17$ (ref. 3), that is, 4.6 mag or 70 times brighter than the derived upper limits. Remarkably, both GRB 050509B, the only short GRB that had previously been localized to better than arcminute precision[8,21], and GRB 050709 were located in low-redshift galaxies[10,21], that is, suitable for a supernova search, but in neither case was evidence for a supernova found in their light curves[22]. Although we cannot entirely rule out the possibility that GRB 050509B originated from a high-redshift star-forming galaxy beyond its presumed host galaxy, this is very unlikely for GRB 050709. Bright SN 1998bw-like core-collapse supernovae are, therefore, effectively ruled out as the progenitors of short GRBs. In the unification (collapsar) scheme, a supernova similar to those seen in long GRBs is predicted[19]. Unless a mechanism to suppress their supernova signals is found, our results disfavour $^{56}$Ni-producing collapsars.

The optical afterglow was located in the outskirts of a star-forming, subluminous host galaxy. Compact object mergers are expected to occur in a more uniformly distributed fashion about their host galaxies than are collapsars, as a result of the evolutionary merging timescale involved (Myr–Gyr)[23]. The compact binary may travel several kiloparsecs before the merger owing to the kick imparted to the binary during the supernova explosions of the progenitor stars. Collapsars occur in starburst galaxies[24], whereas mergers of two neutron stars are expected to have a wide distribution in coalescence timescales[23] and so are expected to occur in both old and young galaxies. We note that the offset of the burst position from its host galaxy centre is among the largest observed[25], and the afterglow is not conspicuously centred on a bright part of the host as is usually seen in long GRBs[26]. Whereas GRB 050509B was associated with an elliptical galaxy, our observations of a starburst host galaxy establishes the diverse nature of these events. This is analogous to the way Type Ia supernovae are found in diverse environments. The different environments and the different mean ages of the stellar population may be indicative of a wide range of formation timescales (for example, the inspiral timescale in the merger model) or simply reflect the likely existence of an old population in starburst galaxies.

At 34 h, the afterglow exhibits a very steep optical to X-ray spectral index, $\beta_{OX}=-1.1\pm0.1$ ($f_\nu \propto \nu^\beta$), steeper than that observed for any afterglow of a long GRB[27]. A merger of two neutron stars may lead to a 'mini-supernova'—that is, emission in the ultraviolet/optical domain at about 1 d after the merger with peak brightness similar to a supernova, and arising from subrelativistic ejecta condensing to form neutron-rich, radioactive material that acts as a heat source for an expanding envelope[28]. In view of the steep optical to X-ray spectral index, our detection of an optical transient may be consistent with a variant of the mini-supernova model; we note, however, that this probably requires the X-ray and optical emission to originate from different emitting regions. Alternatively, the afterglow can be attributed to synchrotron emission from relativistic ejecta[17], much as in long GRBs[12,29] and for the X-ray afterglow of GRB 050509B[8,21,30]. The observed parameters for GRB 050709, $\alpha=-1.33\pm0.45$ and $\beta_{OX}=-1.1\pm0.1$, are consistent with synchrotron emission with a power-law electron energy distribution with index $2.0<p<3.4$. Synchrotron emission from short GRBs is expected to be much fainter than long GRB afterglows as the total energy deposited is expected to be smaller. The much fainter X-ray afterglow of GRB 050509B and the lack of optical detection may be related to its smaller isotropic energy output (~$1.1\times10^{48}$ erg; ref. 8), ~20 times less than in GRB 050709.

The localization of short GRBs have lead to intriguing observations that seem to support the leading merger model[8,21,30]. However, with a sample of only two events it would be prudent not to draw definitive conclusions at this stage about the progenitors of short GRBs. Future optical observations, even from small telescopes, will play an important role in finally determining the origin of short GRBs, although ultimately, detection of gravitational waves[23] will demonstrate whether short GRBs result from mergers of coalescing binary compact stars.

**Acknowledgements** We thank T. Tauris for discussions. The Dark Cosmology Centre is funded by the DNRF. We acknowledge benefits from collaboration within the EU FP5 Research Training Network 'Gamma-ray bursts: an enigma and a tool'.

**Author Information** Reprints and permissions information is available at npg.nature.com/reprintsandpermissions. The authors declare no competing financial interests.

**Correspondence** and requests for materials should be addressed to J.H. (jens@astro.ku.dk).

**Table 1 | Danish 1.5-m optical imaging observations of GRB 050709**

| Date (July 2005 UT) | Band | Int. time (s) | Seeing (FWHM in arcsec) | Afterglow flux (µJy) |
|---|---|---|---|---|
| 11.3587 | R | 12×600 | 0.7 | 2.34±0.12 |
| 12.3283 | R | 17×600 | 1.0 | 1.17±0.26 |
| 17.3233 | R | 24×600 | 1.1 | −0.06±0.23 |
| 18.3999 | R | 13×600 | 0.9 | 0 |
| 19.3991 | V | 3×600 | 1.6 | |
| 19.4062 | I | 2×300 + 2×600 | 1.4 | |
| 19.4106 | B | 3×600 | 1.6 | |
| 27.3799 | R | 12×600 | 1.0 | 0.21±0.17 |
| 29.3957 | R | 8×600 | 1.5 | 0.13±0.24 |

GRB 050709 occurred on July 9.9421 2005 UT. The images were bias-subtracted and flat-fielded in the standard manner. Aperture photometry of the source is complicated by the presence of the bright ($R \approx 21.2$ mag) galaxy, so we used image subtraction[14] to properly measure the afterglow fluxes. The reported fluxes are relative to the R-band image obtained on July 18.4 UT). Extrapolation of the power-law fit to the first data points suggests an afterglow flux at this time of 0.16 µJy, consistent, within errors, with zero flux. In this case the relative fluxes become absolute fluxes as assumed in Fig. 3.

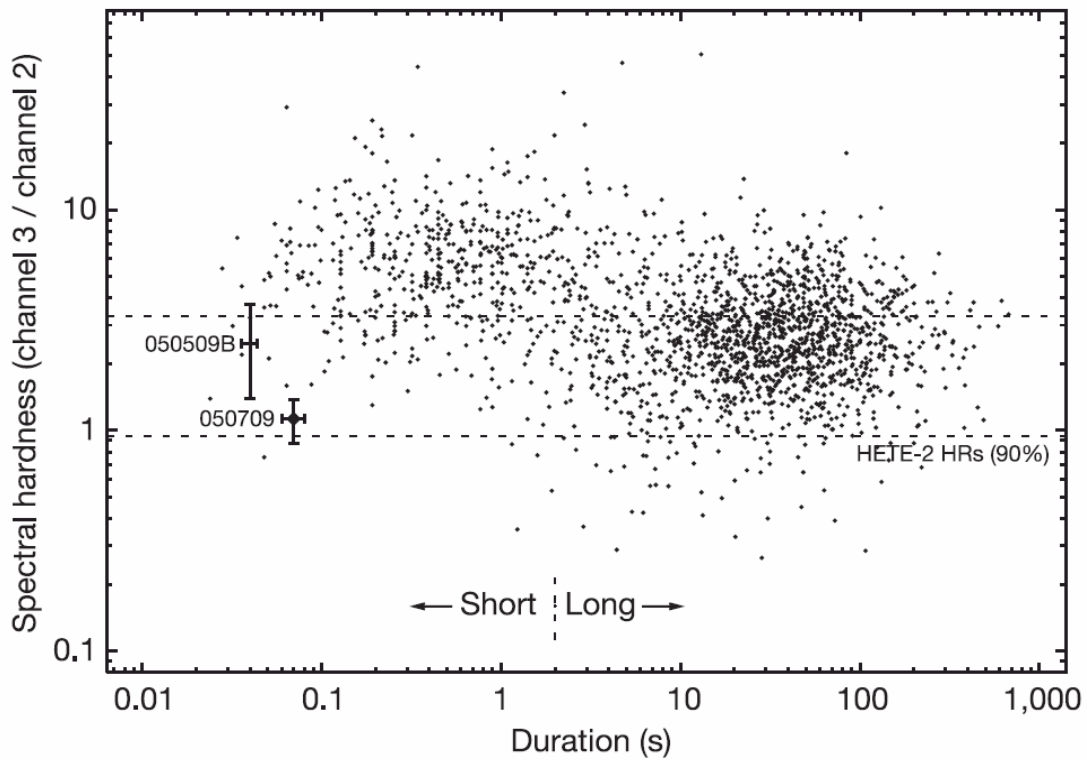

**Figure 1 | The classic BATSE duration–spectral hardness diagram[1].** GRB 050709 (large filled diamond) is clearly among the shortest GRBs, more than an order of magnitude below the canonical 2 s divide between long and short bursts. The burst is also among the softer short GRBs. This fact is probably an instrumental bias effect related to the soft response function of HETE-2's WXM, which is sensitive in the 2–25 keV bandpass (compared to BATSE's nominal trigger range between 50 and 300 keV). Indeed, most long GRBs localized by WXM have soft spectra — the 90% limits for the hardness ratios (HRs) of a sample of 50 long FREGATE/WXM detected GRBs are plotted as dashed lines. The HETE-2 hardness ratios were converted to the BATSE spectral 50–100 keV (channel 2) and 100–300 keV (channel 3) bands assuming a simple power-law model. The only short GRB rapidly localized so far by the Swift-XRT, GRB 050509B, is plotted for comparison. The hardness ratio for GRB 050709 was derived from an exponentially cut-off power-law model fit to the data from the short, hard pulse[9]. (The long, soft component that could not have been detected by BATSE is not considered.) Error bars are 68% confidence limits.

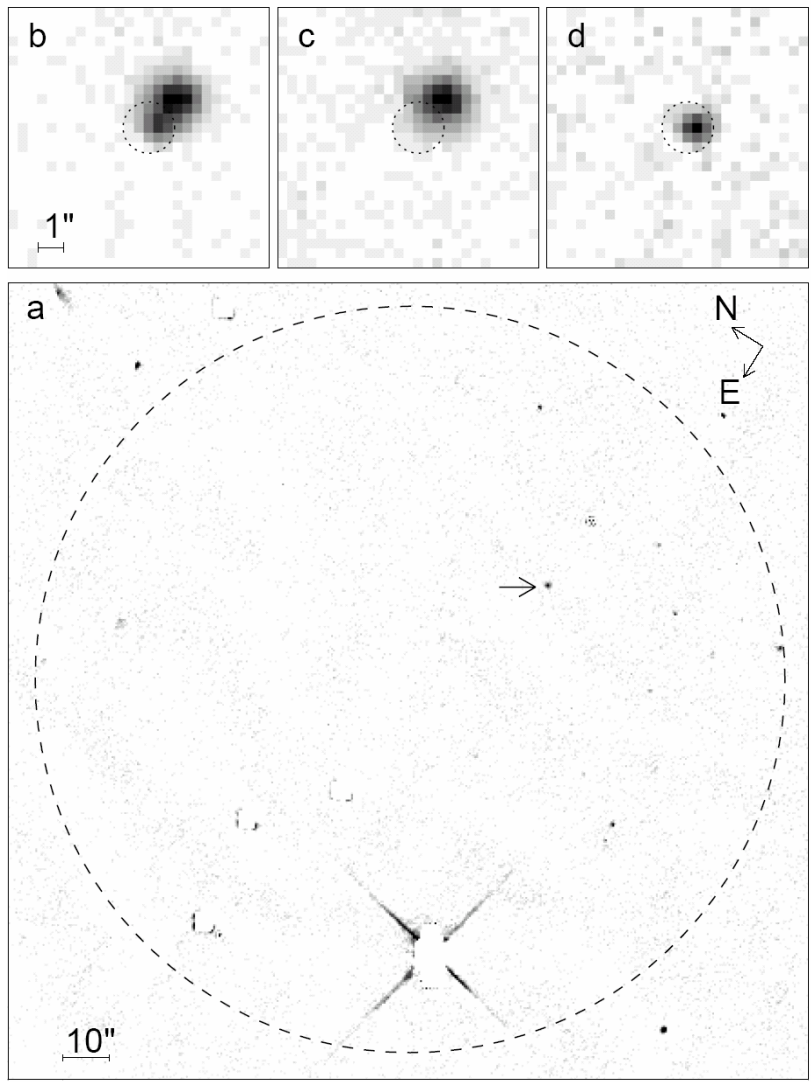

**Figure 2 | The optical afterglow of GRB 050709. a**, Variable sources in an R-band image of the field of GRB 050709 taken with the Danish 1.5-m telescope + DFOSC at La Silla on 11 July 2005, commencing at 6:29 UT. Subtraction of a reference frame taken a week later reveals that only one source inside the indicated HETE-2 SXC error circle[9], identified as the afterglow of GRB 050709, faded away. The subtraction was performed by solving for the best convolution kernel between stars common to the two images[14] using discrete basis functions to describe the kernel. The optical source (arrowed) is situated on the edge of a low-redshift galaxy. **b**, The optical R-band image from July 11.4 UT and the CXO 95% error circle. **c**, The July 18.4 UT R-band image showing the host galaxy after the afterglow has faded away. **d**, The difference between the July 11.4 UT and the July 18.4 UT images showing the afterglow that has faded away over this period. The orientation and scale of the images are indicated.

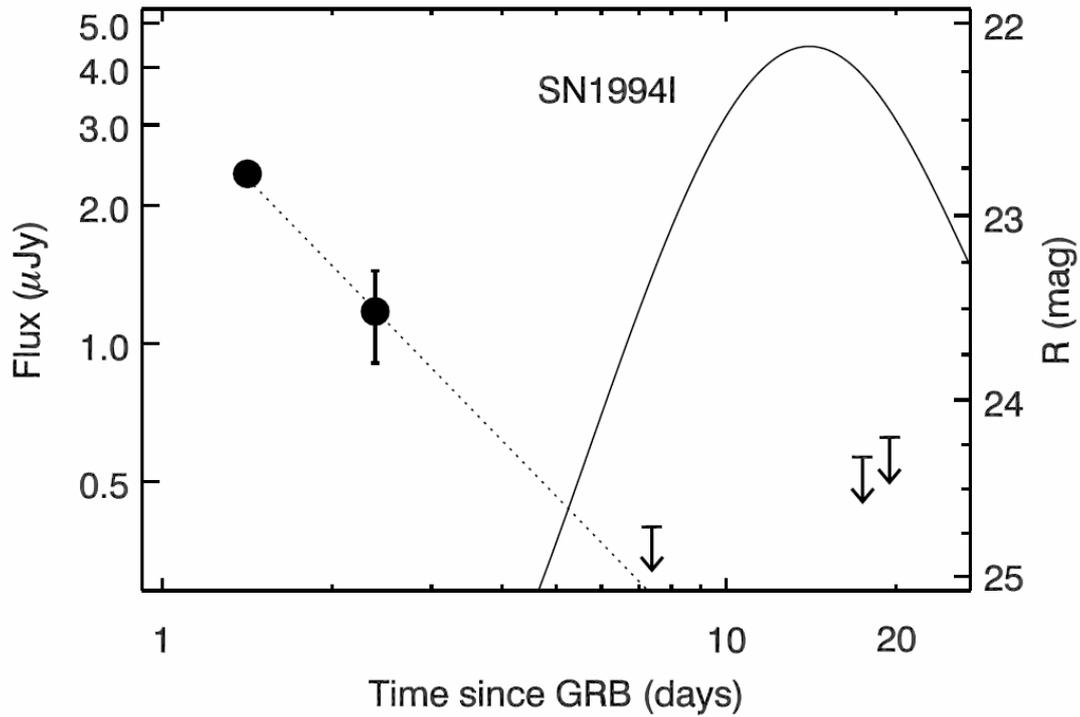

**Figure 3 | Light curve of the optical counterpart to GRB 050709.** The data points are from Table 1. Error bars are $1\sigma$ standard deviations. Upper limits are $2\sigma$ upper limits. Fitting a power law to the first two data points (dotted line) yields a decay slope of $\alpha=-1.33\pm0.45$ (including the first upper limit results in $\alpha=-1.61\pm0.44$). This value is similar to the decay slopes observed for long GRBs, and is indicative of synchrotron emission. Also shown (solid curve) is the light curve of the faint Type Ic SN 1994I, which is fainter than normal template supernovae of types Ia and Ic[22]. Our upper limits at 18 and 20 days clearly rule out even such a faint supernova.